# Liquid-solid Phase Transition Alloy as Reversible and Rapid Molding Bone Cement

**Liting Yi [1], Chao Jin [1], Jing Liu [1,2]***

**1.** Department of Biomedical Engineering, School of Medicine,

Tsinghua University, Beijing, China

**2.** Beijing Key Lab of CryoBiomedical Engineering and Key Lab of Cryogenics,

Technical Institute of Physics and Chemistry,

Chinese Academy of Sciences, Beijing, China

**\*Address for correspondence:**

Dr. Jing Liu

Department of Biomedical Engineering,

School of Medicine,

Tsinghua University,

Beijing 100084, China

E-mail address: jliubme@tsinghua.edu.cn

Tel. +86-10-62794896

Fax: +86-10-82543767

**Abstract**

Bone cement has been demonstrated as an essential restorative material in the orthopedic surgery. However current materials often imply unavoidable drawbacks, such as tissue-cement reaction induced thermal injuries and troublesome revision procedure. Here we proposed an injectable alloy cement to address such problems through its liquid-solid phase transition mechanism. The cement is made of a unique alloy Bi/In/Sn/Zn with a specifically designed low melting point 57. 5 ℃. This property enables its rapid molding into various shapes with high plasticity. Some fundamental characteristics including mechanical strength behaviors and phase transition-induced thermal features have been measured to demonstrate the competence of alloy as unconventional cement with favorable merits. Further biocompatible tests showed that this material could be safely employed in vivo. In addition, experiments also found the alloy cement's capability as an excellent contrast agent for radiation imaging. Particularly, the proposed alloy cement with reversible phase transition feature significantly simplifies the revision of cement and prosthesis. This study opens the way to implement alloy material as bone cement to fulfill diverse clinical needs.

**Introduction**

The diseases in human bone, e.g. osteoporosis often weaken bones physically, thereby leading to the functional disorders. Generally, bones possess the ability of repairing themselves[1], however for many cases, some severely damaged have been clinically verified to be irreparable[2]. In this side, bone cements play extremely important roles in repairing such damaged bones. Among the many traditional materials, polymethylmethacrylate (PMMA), calcium phosphate cement (CPC) and the derivatives have been widely employed in the clinical practices. PMMA has served as very useful cement in spinal surgery and cemented joint replacement[3-5]. However, its exothermically high polymerization in vivo often results in temperatures exceeding 100 ℃[6]. These elevated temperatures can cause irreversible damage of surrounding tissues and further contribute to the failures in joint replacements due to aseptic loosening and volumetric shrinkage[7-12]. CPC, as another essential cement[13,14], was developed relatively quickly owning to its good bioactivity, degradation and osteoconductivity. In spite of this, the problems of the long setting time[15] and easy washout characteristics seriously limit its repairing effect[16]. Notably, the long time for the paste to set into a solid state may induce infection, reducing the operative success rate. In general, radiological imaging has served as an efficient tool to observe and monitor the repairing effect of the cement. However, it is rather difficult to distinguish the traditional cements from the surroundings via radiological images. Efforts have therefore been made to develop cements with additive to increase radiopacity. Previous study has shown that the addition of bismuth salicylate in acrylic bone cements has better effects than barium sulfate[17]. And other additives are also investigated, such as organo-bismuth compound[18], iodine-containing monomers[19], nanosilver and $ZrO_2$[20]. But these methods for increasing radiopacity still involve extra compositions, and the uniform of the cements composites may affect the imaging results.

One of the important applications of cement is in total hip arthroplasty (THA). In recent years, more and more patients are undergoing THA, which is proved to be a highly successful operation in relieving pain and recovering function. Even so, a proportion of failures are inevitable. Thus, the rates of revision of THA have been increasing[21,22]. Clearly, removal of a failed THA can be a big challenge and risk damage to the remaining bone stock. Usually, it requires specialized manual instruments and technical skills for surgeons[23,24]. The commonly adopted means for removing prosthesis and cement include the sliding trochanteric osteotomy, the extended trochanteric osteotomy (ETO), the extended proximal femur and the cortical windows[25-30]. Clearly, these methods require surgeons to cut femurs, such as the ETO that it involves the removal of 1/3 of the femoral shaft[29], which not only weakens the remaining bone but also is associated with complications, longer operating time, and greater blood loss compared to primary total hip replacements[31].

An appropriate bone cement for repairing vertebral fracture is characterized by proper injectability, rapid setting, adequate stiffness, bioactivity, low setting temperature, and radiopacity[15]. Given these requirements, we are dedicated here to propose a strategically different approach to make bone cement with predominant properties to overcome the limitations of traditional materials. In fact, a long-time study has been launched on metal as biomaterials used in orthopedics, such that stainless steel, cobalt alloy, titanium and titanium alloy are most commonly used[32-34]. However, such metal materials generally have high melting point and low plasticity which would meet trouble in forming various shapes in situ. From another alternative, we proposed the low melting point metal as bone cement. As is noted, the eutectic alloy which was first reported as a suitable solder has the melting temperature as low as 60 ℃[35]. In addition, the biocompatibility of similar liquid alloy has been investigated[36]. As metal alloy, it demonstrates certain strength and possesses low melting point and reversible physical phase transition capability. All these characteristics remind us that such eutectic alloy can be used as high quality bone cements.

In this study, we demonstrate a liquid-solid phase transition alloy cement composing of Bi/In/Sn/Zn alloy. Compared to the conventional cements, this new material features simple preparation and operation, rapid setting up, low peak temperature increase and strong radiopacity. We find that this material can be easily shaped into various forms at low temperature to fit for various shapes of defects. Its mechanical strength and in vitro cytotoxicity were determined and clarified. Importantly, this cement allows the reversion much easier and controllable, which is a key to prevent remaining bones from more damages.

**Results**

**Plasticity of liquid-solid phase transition alloy cement**

In order to determine the plasticity of the liquid-solid phase transition alloy cement, molds with letter shapes were employed. Due to the solid state at room temperature, such alloy cement was melted by heating it above the eutectic temperature for preparation of filling into molds. With the aid of injectability and rapid solidification, the alloy cement could be fabricated with ease in various forms. The molds were eliminated until the cement completely becomes solidified. The result shows that the word of "BONE" consisting of four letters made of alloy cement is clearly formed (see Fig. 1a). The whole process took about only 10 min from injection to molding. Taking into account the relationship between plasticity and flow characteristics, the injectability of cement was displayed indirectly via droplet experiments. The dropping process was recorded by a high speed camera (see Fig. 1b). The liquid alloy cement was prepared by heating the solid cement until melted and suction was applied by retracting the plunger of the syringe. Under the room temperature, the liquid alloy cement was vertically pushed out of the syringe in one droplet. The droplet fell to the ground and gradually formed into the liquid film. This process was similar to that of water, differently, the solidification of alloy happened after liquid film formed. We set the time of droplet contacting with ground as 0 ms. The liquid alloy spread fully on the plane in 8 ms, and completed the fluctuation and solidification in 28 ms. Through molding and droplet experiment, we can find that the solidification time significantly depends on the volume, and in inverse proportion. The liquid alloy cement with certain volume will not solidify immediately at room temperature, which allows the operable time for surgeons.

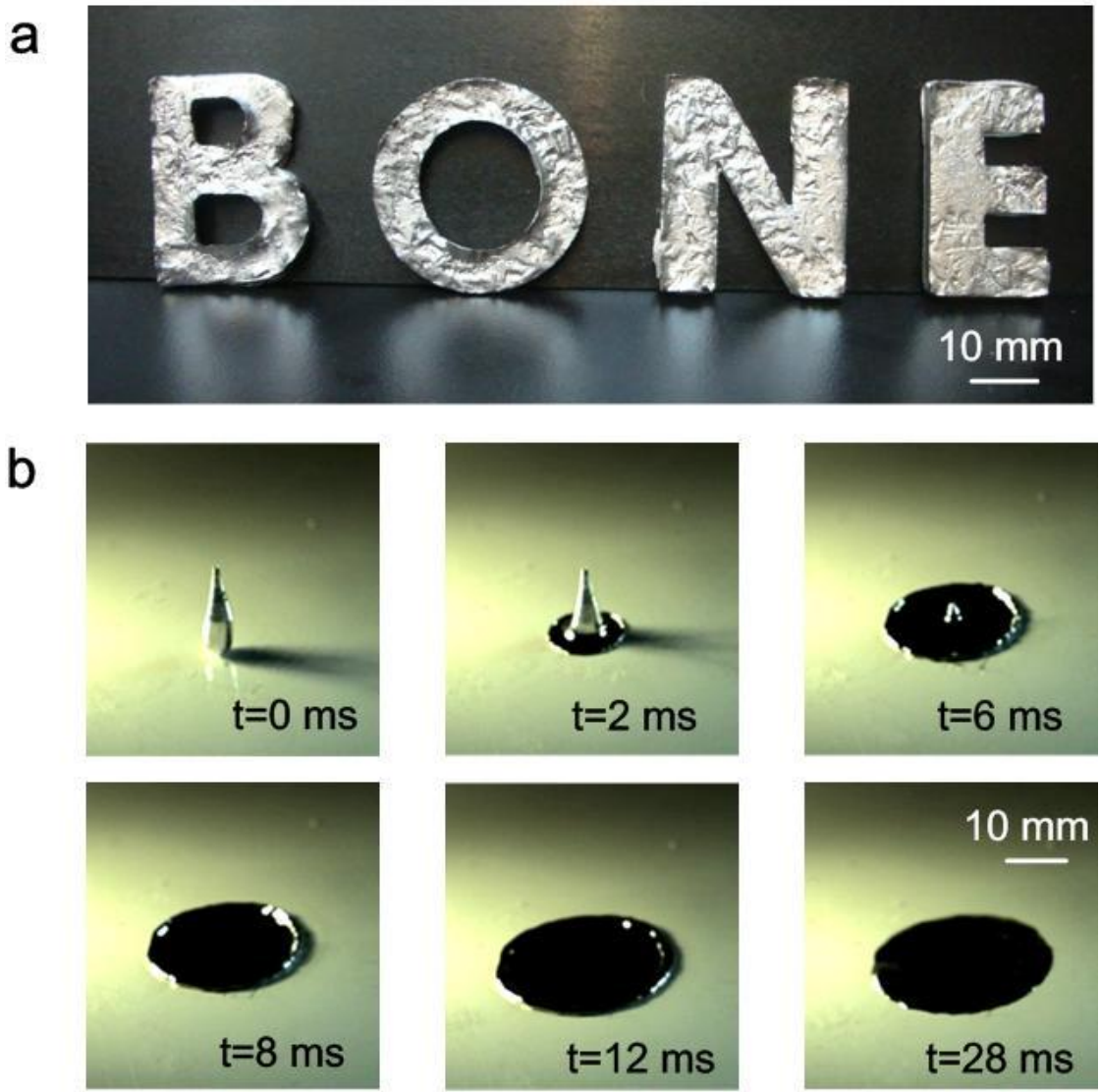

**Figure 1 | Plasticity characterization of liquid-solid phase transition-based alloy cement.** (**a**) A demonstrated molding of alloy cement in a shape of "BONE". (**b**) The transient liquid-solid phase transition process of liquid alloy droplet captured by the high-speed camera.

**Mechanical characteristics**

Proper mechanical characteristics are the basic requirements of a bone cement. By using a universal testing machine, we have studied three mechanical properties (bending strength, compression strength and Young's modulus) of Bi/In/Sn/Zn alloy cement (see Fig. 2). The cylindrical specimens utilized were with diameter of 15 mm (see Fig. 2a and Fig. 2c). The long ones with length of 140 mm (see Fig. 2a) and the short ones (see Fig. 2c) with length of 10 mm were used for bending and compression tests, respectively. The implement methods are shown in Fig. 2b and Fig. 2d (see the details in Methods section). In the three-point bending test, the force was exerted on the middle of the specimen at length and it caused deformations and displacements. The displacements and relative force magnitudes ($F_1$) were recorded (see Fig. 2e). The bending strength ($\sigma_{bend}$) was calculated by the following equation:

$$\sigma_{bend} = \frac{8L_1 F_1}{\pi d^3} \quad (1)$$

where $L_1$ is the support span, 80 mm; $d$ is the diameter of specimen. With respect to the compression test, the deformations and displacements were induced by the force ($F_2$) acting on the top surface of specimens (see Fig. 2d). Equation (2) was utilized to obtain the compression strength ($P$):

$$P = \frac{4F_2}{\pi d^2} \quad (2)$$

Besides, Young's modulus ($E$) was derived from the stress-strain curves, which can be estimated by the equation (3):

$$E = \frac{P}{\Delta L / L} \quad (3)$$

where $L$ is the length of the short specimen, 10 mm; $\Delta L$ is the length variation. The results are shown in Fig. 2e, 2f and 2g, respectively. The measured magnitudes of the compressive strength, bending stress and elastic modulus of the alloy cement were 20 MPa, 44 MPa and 1.6 GPa respectively.

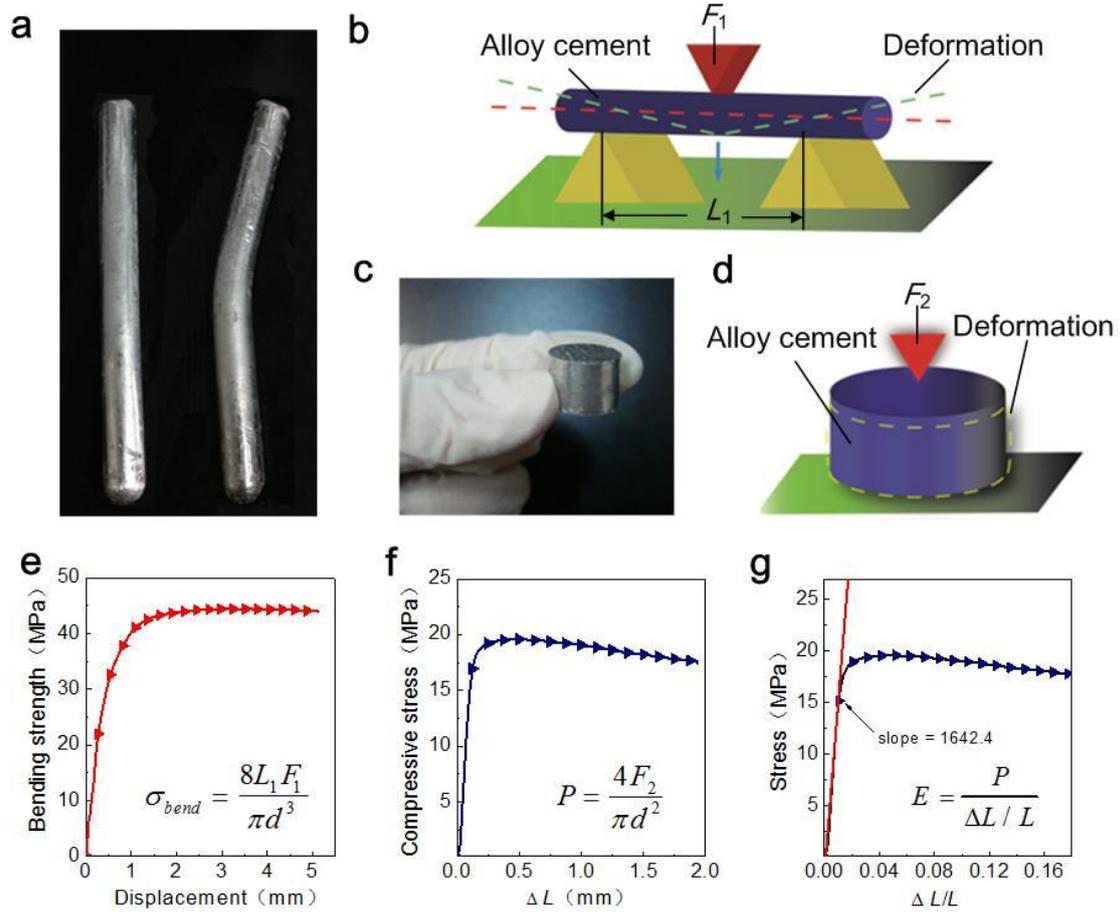

**Figure 2 | Mechanical strength tests of liquid-solid phase transition alloy bone cement.** (**a**) The optical image of cylindrical alloy specimens for bending tests; Here the left and right objects are respectively the same specimen before and after the bending test. The corresponding schematics of bending and compression tests are demonstrated in (**b**) and (**d**), respectively. (**c**) Optical image of the specimen for compression test. The measured bending strength curve (**e**), compression strength curve (**f**) and Young's modulus curve (**g**).

**Thermal physical properties**

Thermal physical characteristics of cement are critical parameters in the in vivo application.

To investigate the clinical potential of alloy cement, we have employed DSC (differential scanning calorimeter) method to assess its fundamental properties. The alloy cement weighted of 29 mg was added into $Al_2O_3$ crucible. The prepared sample was then put into DSC and implemented by the predetermined dynamic program. During this process, the temperature first increased to 100 ℃ at 10 ℃/min, then maintained at isothermal phase for 5 min, finally decreased to -20 ℃ at 10 ℃/min. The DSC curve is demonstrated in Fig. 3a which shows the melting point of Bi/In/Sn/Zn alloy with 57.5 ℃ and the peak rate of heat release with 0.75365 mw/mg.

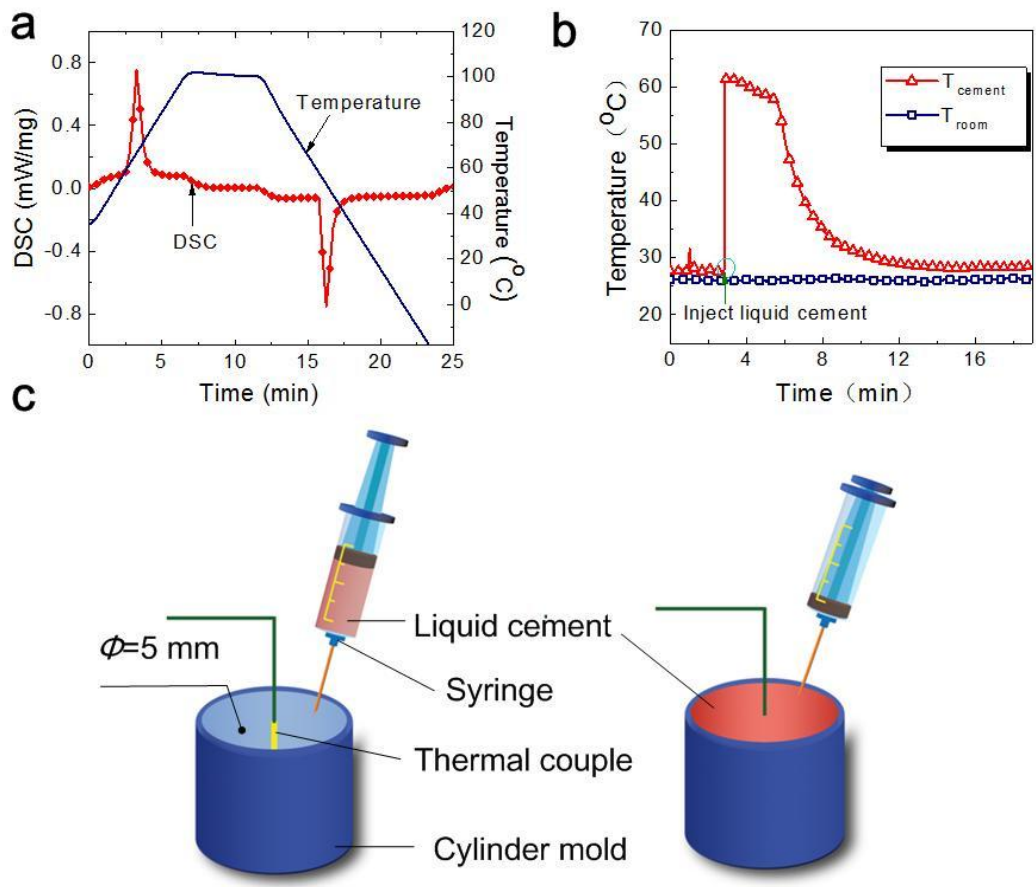

**Figure 3 | Thermal physical properties of the alloy cement.** (**a**) The DSC curve of Bi/In/Sn/Zn alloy (weight percentage, Bi 35%, In 48.6%, Sn 16% and Zn 0.4%). (**b**) The central temperature curve of cylindrical mold during the alloy cement injection process. The temperature rapidly rose to peak value at around 60 ℃ with the cement injection and then decreased slowly in several

minutes. (**c**) Schematic of the peak temperature measurement.

With respect to a new cement, it is significant to present with peak temperature (maximum temperature during the exothermic process, is obtained from the maximum of the exotherm curve). We have utilized the classical method to determine the peak temperature in the liquid-solid transition process. The operational method is shown in Fig. 3c. The thermal couple was placed in the center of the mold for recording the variation of the temperature. The temperature of the alloy cement reached maximum of 61.9 ℃ after the cement was injected into cavity of the mold (see Fig. 3b). Subsequently, the central temperature decreased dramatically within a few minutes. This peak temperature is different from that measured by DSC, because liquid alloy cement injected was preheated at a higher temperature. Thus, the peak temperature may be a little higher than the melting point and closer to the heating temperature. Nevertheless, the peak temperature can be close to the melting point owing to the controllable heating temperature, and it is still a lower temperature compared to traditional cements. Besides, setting temperature is another important parameter related with thermal physical characteristics. It is defined as the period between the start of mixing and the time when the temperature had reached midway from the ambient (20 ℃) to peak temperatures from the recordings[37-39]. And it has been indicated that the lower setting temperature would bring about the better effects in surgery. The method shown in Fig. 3c was also adopted for measurement of setting time. For alloy cement, the central temperature increased immediately to the peak value since the cement was injected to the mold. So its setting temperature does not exist or can be regarded as zero which is superior to that of chemically-reduced phase transition.

**Biocompatibility demonstrations**

The safety of the present cement materials within the human body is a common concern to the future application. To clarify such issue, we have conducted the cytotoxicity tests to demonstrate its biocompatibility by using the relative growth rate (RGR) of the tested cells. The RGR represents the relative cell viability which is calculated by the ratio of OD (optical density) value between the test group and the control group. The relationship between the RGR and cytotoxicity is as follows: RGR >100%, 75-99%, 50-74%, 24-49%, 1-25% and 0% respectively indicate class 0, 1, 2, 3, 4 and 5. The class 5 denotes the highest toxicity, while the class 0 represents no toxicity[15].

In this study, BALB/c 3T3 cells are chosen as objects under investigation and divided into two groups: the test groups treated with 100% cement extract; and the negative control group cultured simply with culture medium. After incubation, the NRU (natural red uptake) was measured at 540 nm and read out in OD value by using a microtitre plate reader (see Fig. 4 and the method section for the details about the cytotoxicity tests). The OD value for control and test groups are 0.1189 ± 0.0211 and 0.1209 ± 0.0420 respectively. Thus, the RGR for the 100% cement extract is 101.7% ± 35.3%, clearly, the corresponding cytotoxicity is between class 0 and 1. It indicates that such alloy cement has no cytotoxicity; specifically, the alloy cement has fewer effects on the normal proliferation of 3T3 cells and the mitochondrial enzymes.

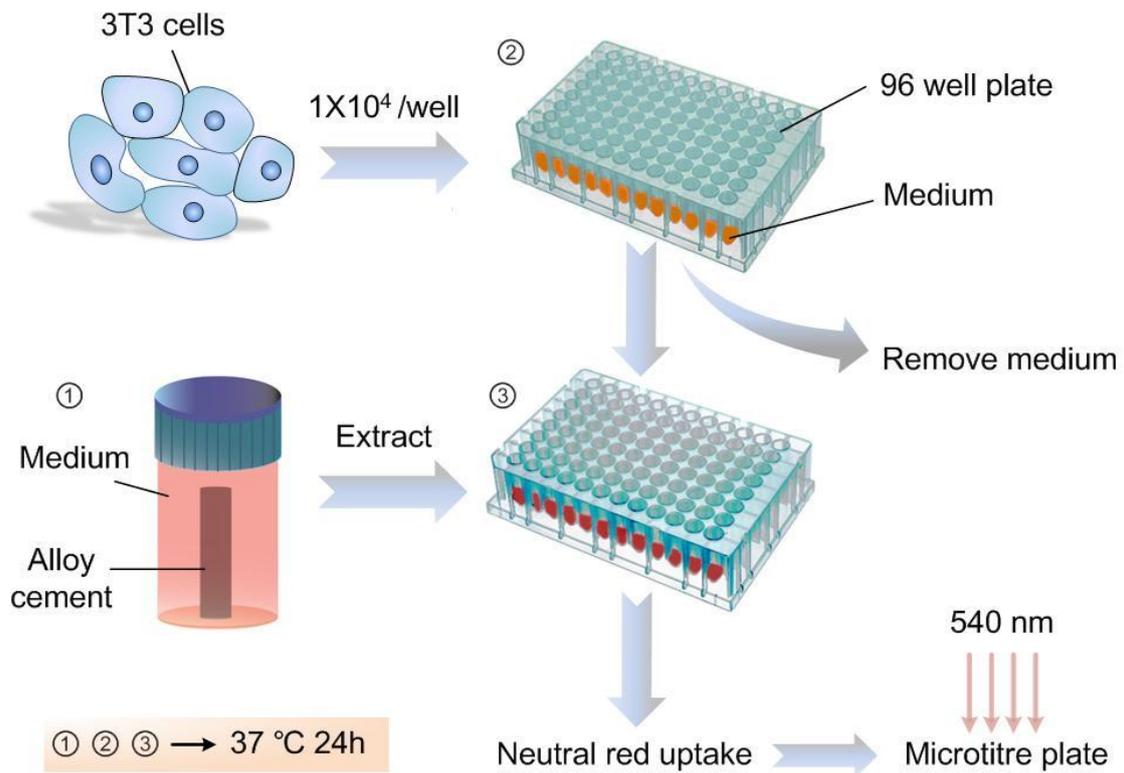

**Figure 4 | Schematic illustrations for the cytotoxicity tests of liquid-solid phase transition alloy cement.** First, the alloy cement was shaped in cylinder for preparing specimen and extracted in culture medium. Meanwhile, 100 μl of BALB/c 3T3 cell suspension of $1 \times 10^5$ cells/ml was seeded into each of the 96 well plates. After incubation, replace the culture medium with cement extract. The NR absorption was then detected at 540 nm. The extraction and incubation process involved were all under the condition of 37 °C for 24 h.

**Contrast agent for radiation imaging**

The radiation imaging is an effective way to monitor and evaluate the status of molding cement in vivo clinically. To research the imaging contrast characteristic of liquid-solid phase transition alloy cement, the radiation images of porcine femurs were taken by X-ray imaging and CT (computed tomography) (see Fig. 5). For comparison, the femur we used is shown in Fig. 5a, and it was reconstructed by CT which demonstrated the actual situation of femur sample. We can clearly observe that such alloy cement show excellent imaging ability with high contrast in X-ray photo before and after cement injection (see Fig. 5b, 5c). This imaging advantage greatly contributes to the cement monitoring after surgery.

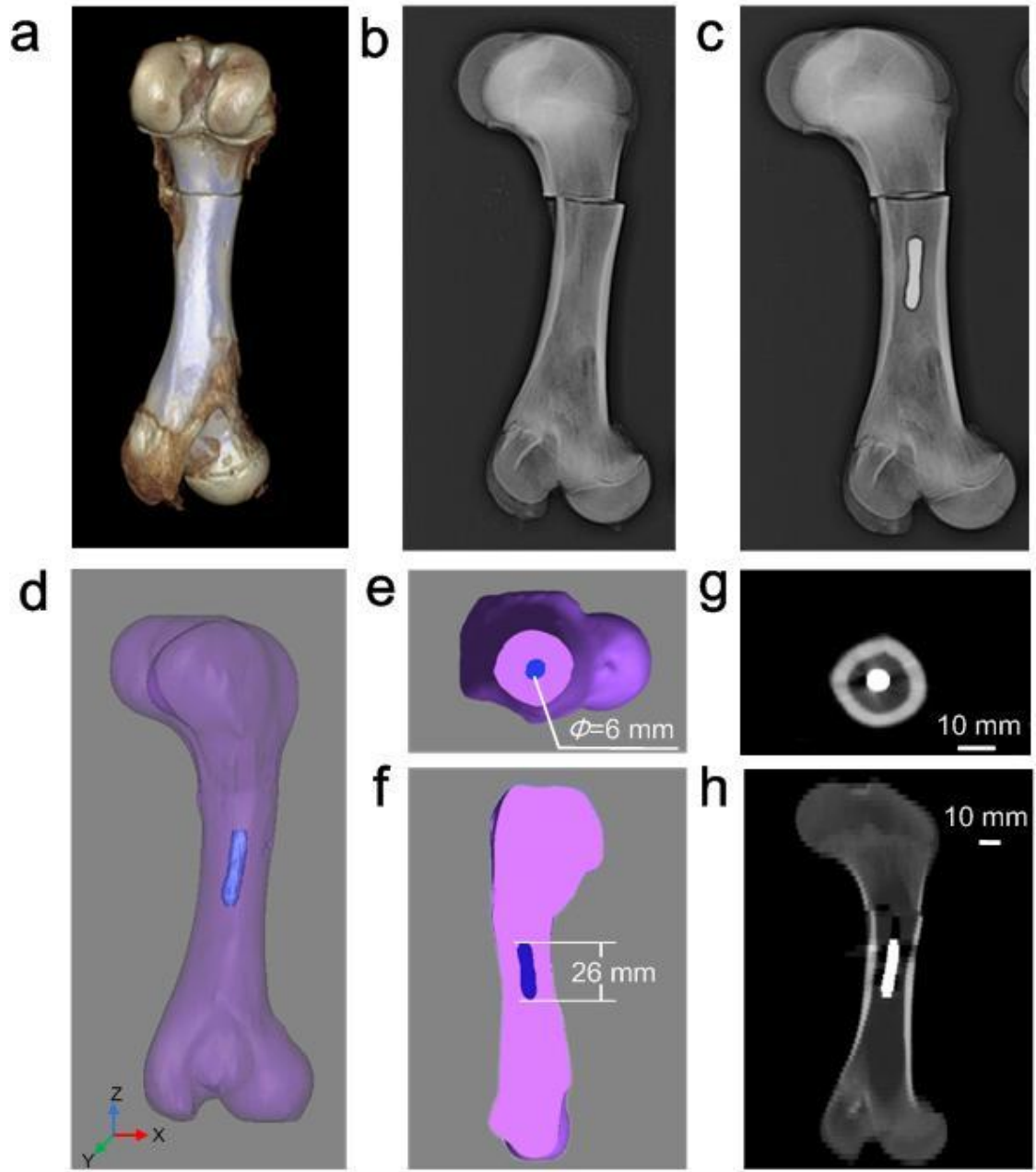

**Figure 5 | Radiation imaging of porcine femur with alloy cement.** (**a**) The reconstructive femur image of CT. (**b**) X-ray imaging of bone with partial excavation of marrow. (**c**) X-ray imaging of

the cavity in femur filled with alloy cement. **(d), (e)** and **(f)** demonstrate the reconstructed geometrical model. (**d**) The geometrical model reconstructed by the CT slices. (**e**), (**f**) The horizontal and coronal cross sections of geometrical model. (**g**), (**h**) The transverse and coronal sections of CT slice, respectively.

Although the X-ray photo is the main imaging method in bone diseases, it simply reflects the 2-D (dimension) information which sometimes is insufficient to master the whole structure of cement filling quality. So, the CT was carried out and the results were reconstructed subsequently (see Fig. 5g, 5h). Owing to the great performance in X-ray imaging, the imaging efficiency of this alloy cement through CT has been greatly improved. Generally, the artifacts of metal in CT severely decrease the image quality. But the results are different here and it shows clear edges of alloy cement (see Fig. 5g, 5h) which extends its imaging role for detection. It also demonstrates that the hole of marrow and the alloy cement were in better shape matching. In addition, the shape of the filling cement could be measured from geometrical model reconstructed by the CT slides (Fig. 5e, 5f). The obtained quantitative information offers high contrast reference for medical measurement.

**The reversible capabilities of the alloy cement for smart surgical revision**

To verify that the solid alloy cement can be easily removed, we used a cylindrical mold to simulate bone in vivo (see Fig. 6a) and implement the revision process. The constant heating temperature of probe was chosen at 200 ℃. The real-time temperature in periphery was recorded by thermal couple at 1 s intervals and the result of such transient temperature profile was shown in

Fig. 6b. Based on the large heat conductivity of alloy cement, when the heating probe contacted its surface, the heat transferred immediately from center to surroundings. The peripheric temperature was increased to 60 ℃ within 1 min. Meanwhile, the solid cement in mold had melted completely. In this case, to avoid overheating, we can remove the heating probe, then aspirate the liquid alloy cement by syringe. The Fig. 6b shows that the operating time for aspirating could be maintained for 5 min. This won more time for surgeon to remove the melting cement. Based on this, the extraction of prosthesis was also easily accessible.

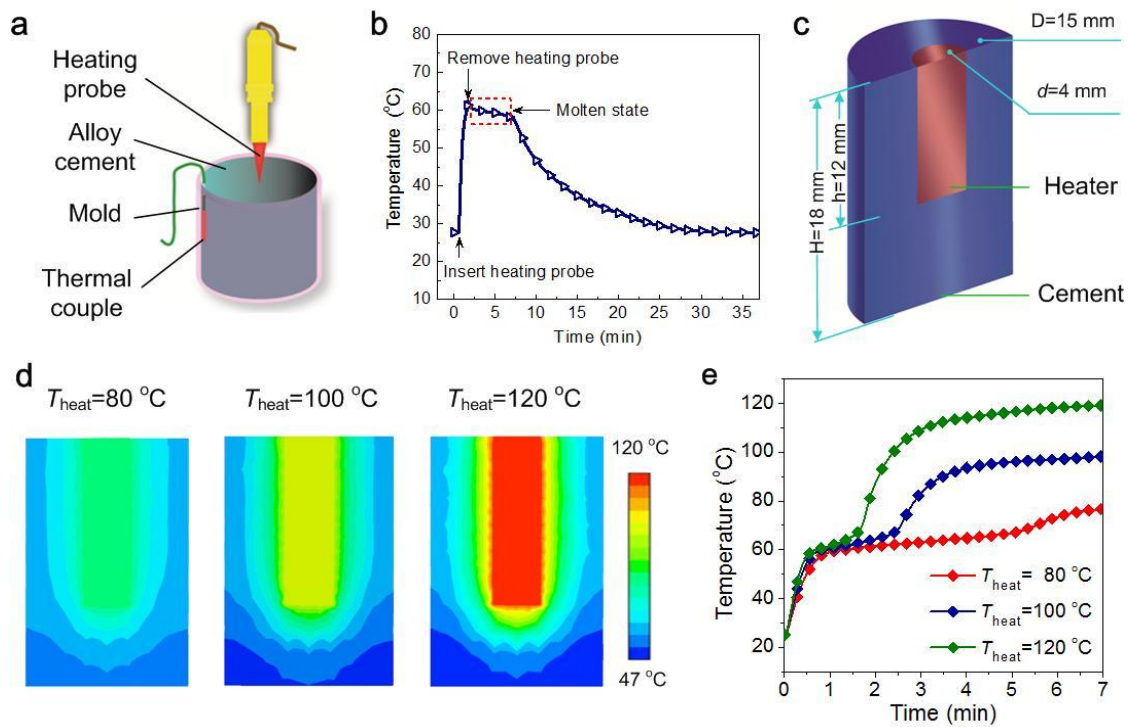

**Figure 6 | Thermal features of the alloy cement during the solid-liquid phase transitions.** (**a**) Schematic of heating solid alloy cement. The heating probe was placed on the center of cement, and the thermal couple was used to monitor the temperature in periphery. The temperature during the operation process was recorded (**b**). The cement could be easily removed via syringe when it was melting. (**c**) The geometrical model for simulating the thermal effects during the heating. (**d**)

The simulated temperature distributions on the cross-sectional plane in the condition of $T_{heat}$= 80 $^{o}$C, 100 $^{o}$C, 120 $^{o}$C; Here, $T_{heat}$ denotes the heating temperature. (**e**) The real-time maximum temperature profiles on the sidewalls in three cases of heating conditions.

We inferred that the required time of alloy for melting had strong correlation with the heating temperature. To substantiate this, a computational model focusing on relationship between heating temperature and thermal distribution was established for further analysis. In this model, three conditions of central heating temperature ($T_{heat}$=80 ℃, 100 ℃ and 120 ℃) were set respectively, and other parameters kept constant (see Supplementary). Fig. 6d shows that the temperature distributions when peripheric temperature was increased to 60 ℃ under different heating conditions. The temperatures increased since the heating source was inserted, but they were lower with the displacement increasing. Besides, the variation of temperature on the edges was simulated in the case that the heating probe kept working for 30 min. This investigation takes into account the injury risk from the delay of removing heating probe resulting in operational errors. The heating rate was positively proportional to the heating temperature, so it took longer time for lower heating temperature to achieve all cement melting but the low temperature may reduce the risk of thermal injury (see Fig. 6e). In addition, the shape of heating probe is also essential for different shapes of the filled cements. Thus, the concrete parameters of heating probe could be designed and determined by practical demands.

**Discussion**

Current cements are generally composed of powder and liquid systems. Prior to use, these two systems are mixed in situ at an appropriate ratio. But for alloy cement, it only has one system that solid and liquid phase can be transformed flexibly, thereby mitigates the complexity of the preparation work. During the solidification of alloy cement, because of the lower ambience temperature, the temperature decreases continuously and rapidly from the peak which is approximate to its melting point. The low peak temperature may decrease the incidence of premature failure of the cement mantle, which improves the effects in repairing[40]; while the rapid molding can reduce the chances of infection in the surgery. Besides, the measured mechanical strength of this alloy material is close to CPC but slightly lower than PMMA. Nevertheless, alloy cement could be employed as potential cement material, particularly in the repairing of non-weight-bearing bones, e.g. craniofacial reconstruction. Meanwhile, our further studies would attempt to improve the mechanical strength of such cement by adding nanoparticles or other metals to modify them to be more specifically effectively and widely adopted. The NRU cytotoxicity test shows that the alloy cement has very low or even no toxicity for 3T3 cells. These results indicate that this material is very possible to be used in vivo, and animal experiments will be carried out to further confirm the quality. Here, the alloy cement itself also plays a role of contrast agent which is necessary for operation monitoring. Significantly, there is an evident advantage of alloy cement in revision owing to its reversible properties in liquid-solid phase transition. Further, the heating means involved are not only probe, but also direct heating on prosthesis. The materials of prosthesis are generally metal with good thermal-conductivity. Therefore, for cemented prosthesis, the cement attached to the prosthesis melts when prosthesis is heated, which results in the reducing of resistance on cement-prosthsis surface. Thus, the cement and prosthesis can be easily removed.

We have developed a liquid-solid phase transition cement consisting of Bi/In/Sn/Zn alloy. The results suggest that this alloy cement shows excellent characteristics of reversible, rapid setting, lower peak temperature, radiopacity, and low cytotoxicity. Its reversible property diminishes the trauma of surgery and reduces the complexity of operation. As a kind of metal material, alloy cement is recyclable which greatly saves the cost. Alloy materials also have good performance in electrical and thermal conductivity. It has been proved that the electrical stimulation is of great importance in controlling bone growth and healing[41,42], and the electrical polarization of titanium implants may increase the osteoblast differentiation[43]. For cemented component, the electrical stimulation of traditional cement layer is restricted with the poor electrical conductivity. In contrast, the excellent performance of our alloy cement in electrical conductivity has shown potential to be effective in aiding bone healing. Moreover, when they are applied in bone defects resulting from bone tumors, the heating alloy cements are capable of killing surrounding residual cancer cells by hyperthermia. Although further investigations are needed, the liquid-solid phase transition alloy cement has shown great promise as novel bone cement with improved properties and it will change the basic concept of cement material.

**Materials and method**

**Preparation of cement**

The liquid-solid phase transition alloy bone cement was composed of bismuth (Bi), indium (In), stannum (Sn) and zinc (Zn) metals with purity of 99.99 percent, and they were weighted with a corresponding weight ratio of 35%, 48.6%, 16% and 0.4%. The weighted metals were added into the beaker and heated at 400 ℃ for 24 h to prepare alloy. To increase the homogeneity of the components, the mixture were stirred using a magnetic stirrer at 70 ℃ for 4 h when they were all melted. Finally, the obtained alloy cement with melting point of 57.5 ℃ could be stored at room temperature in solid state for a long time. Before the injection, the alloy cement only needs to be heated at the temperature higher than 57.5 ℃. The device utilized for injection was an ordinary syringe without needle.

**Plasticity evaluation of liquid-solid phase transition alloy cement**

Firstly, the flow characteristics of Bi/In/Sn/Zn alloy were demonstrated by studying the dropping process. After melting, liquid alloy cement was injected by 1 ml syringe, and the piston was forced to form droplet at room temperature. The metal droplet was formed vertically at room temperature. The high-speed camera (NR4-S3) was employed to record the dropping process at 500 fps (frames per second). To evaluate the plasticity, we injected the melting alloy cement into different letter molds. The molds with height of 3.5 cm were placed on the heating platform with 70 ℃ to keep the cement in liquid state during the filling process. After the molds were full of cement, remove them from the platform and leave them at room temperature. About cooling for 10 min, the molding alloy was released from the molds.

**Mechanical characteristics evaluation**

Cylindrical specimens (15 mm in diameter, 10 mm in length) were prepared for compression testing, and other cylindrical specimens (15 mm in diameter, 140 mm in length) were prepared to determine the bending strength by injecting the melting cement into the molds. The values for Young's modulus was derived from the stress-strain curves measured from the compression tests.

The specimens were cured within molds at room temperature for 24 h. The mechanical test was performed after the cement became hardened. Three specimens of cement were used for each test. The values of ultimate compressive strength and bending strength (a three-point bending test) were measured using a universal testing machine (Autograph Model AG-X, Shimadzu, Kyoto, Japan) at a cross-head speed of 0.5 mm/min.

**Thermal physical properties evaluation**

The thermal property of Bi/In/Sn/Zn was measured by DSC (NETZSH, 200 PC). Scan was implemented in the following process: the temperature first increased to 100 °C at 10 °C/min, then maintained at isothermal phase for 5 min, finally decreased to -20 °C at 10 °C/min. Moreover, to determine the peak temperature during solidification, the PP (Propene Polymer) cylindrical molds (5 mm in diameter of inner bore) were employed. The molds were placed at room temperature. A thermocouple was inserted into the center of the mold, and the temperature was measured at 1 s intervals. Subsequently, the liquid cement was injected into the mold. And the temperature was recorded until it drops down to the environmental temperature.

**Biocompatibility evaluation**

Instead of culture medium, the cement extract was used to culture BALB/c 3T3 cells. The culture medium for 3T3 cells was DMEM supplemented with 10% fetal calf serum and 1% Penicillin-Streptomycin solution. The cement was shaped in cylinder for preparing specimens and extracted in culture medium (3 cm$^2$/ mL) according to the standard International Standards Organization (ISO) 10993-12 for 120 h at 37 °C[44]. The Netural red uptake (NRU) cytotoxicity was tested according to the standard International Standards Organization (ISO) 10993-5[45]. Cell suspension of $1\times10^5$ cells/mL was added into 96-well tissue microtitre plate with culture medium. After 24 h incubation, replace the medium with 100 μL cement extract (100% extract) for another 24 h incubation (5% $CO_2$, 37 ℃), except that the negative control was replaced with culture medium. Then the cells were processed by 100 μL Neutral red (NR) medium for 3h. Subsequently, remove the NR medium, wash once with 150 μL PBS and add 150 μL NR desorbing fixative. The OD value was measured at 540 nm in a microtitre plate reader (SpectraMax M5), using the blanks as a reference.

**Contrast agent for radiation imaging**

Four fresh porcine femurs for X-ray imaging (General Electrics, XR/A) and two for CT (Philips Brilliance 6) were used to verify the radiopacity of the alloy cement. Femurs were cleaned of excess soft tissue and stored for a short period of time in 4 ℃ before use. The entire bones were cut into two aspects to expose the marrow cavity and excavate the marrow partially for the injection of liquid alloy cement. Set the femurs in thermostat water bath (37 ℃) or at room temperature, then inject the melting cement into the marrow cavity using 1 mL injection syringe (without needle). When the naturally cooled cement was converted into solid state, the radiographs were taken. Besides, the geometrical model was reconstructed by the CT slices.

**The reversible features of the alloy cement**

The melting alloy cement was placed in a PP (Propene Polymer) cylindrical mold (15 mm diameter and 18 mm height). The thermal couple was inserted in the alloy cement closing to inner surface of the mold. After the specimen of alloy cement became hardened and its temperature reached the ambient condition, the solid alloy cement was heated until melted completely by the heating probe with temperature of 200 ℃. In the meanwhile, the temperature was recorded through thermal couple with time interval of 1 s (Agilent 34907A). The working part of the probe was conical with diameter of 6 mm and height of 12 mm and it was kept placed in the center of the specimen. Along with the alloy melting, the probe was moved down vertically. Once the specimen became completely melted, remove the heating probe immediately.

In addition, a computational model was established to further analyze the correlation between heating temperature and heating rate. The detailed dimensions of model were shown in Fig. 6c, and both of the alloy cement and heating probe were cylinders. Under three conditions ($T_{heat}$=80 ℃, 100 ℃ and 120 ℃), the temperature distributions and curves were simulated and evaluated.

## Appendix:

## Theoretical simulations on the thermal distributions

### Theoretical model

Theoretical model was used to simulate the thermal distribution of alloy cement during the heating process. Fig. 6c provides the detailed dimensions of theoretical model. Here, the classical heat transfer equation was utilized to characterize the transient 3D temperature distribution in the alloy cement:

$$\rho c \frac{\partial T(\mathbf{X},t)}{\partial t} = \nabla(k \nabla T(\mathbf{X},t)) \tag{1}$$

where $\mathbf{X}$ denotes the Cartesian coordinates $x$, $y$ and $z$; $\rho$ denotes the density of Bi/In/Sn/Zn alloy, 7.564 g/cm$^3$; $k$ is the thermal conductivity, 13 W/(m ℃); $c$ is the specific heat which transiently varies with temperature. To implement more accurate simulations, we fit the curve of $c$ by B-spline interpolation function and the estimated result is shown in Fig. S1.

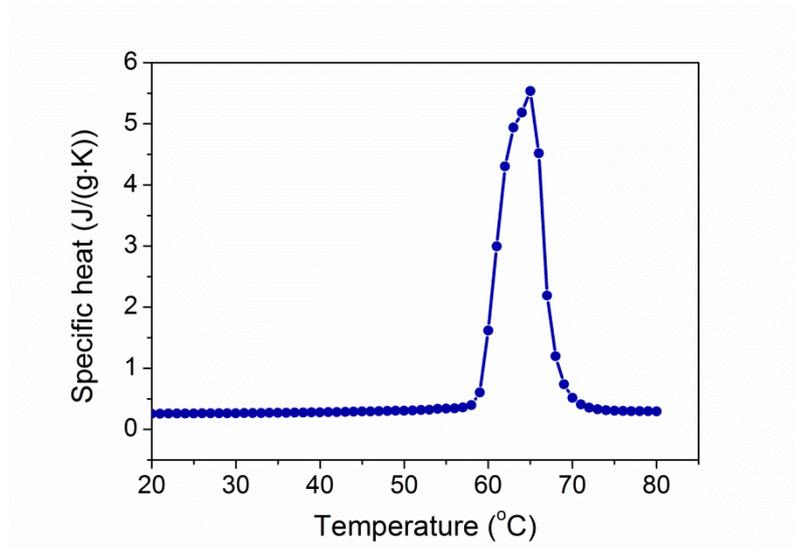

**Figure S1 | The specific heat curve of Bi/In/Sn/Zn alloy via fitting.**

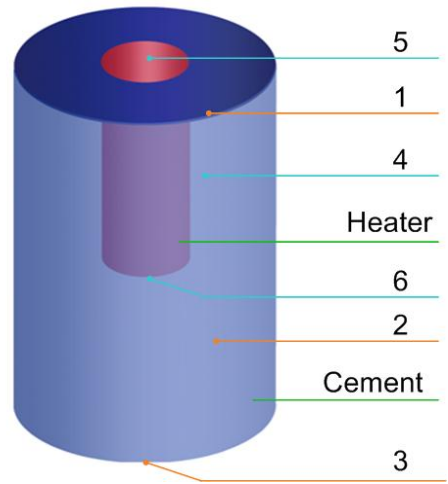

**Figure S2 | Geometric theoretical model for thermal distribution.** 1-6 are for the boundary conditions of theoretical model. 1-3 denote the upper, profile and bottom surface of the alloy cement. 4-6 denote the profile, upper, and bottom surface of the heating probe.

The corresponding thermal boundary conditions of 1-3 and 5 (see Fig. S2) are expressed in equation (2). And the equation (3) gives the boundary conditions of 4 and 6.

$$k \frac{\partial T}{\partial n} = 0 \qquad (2)$$

$$T = T_{heat} \tag{3}$$

The heating temperatures $T_{heat}$ under three conditions were 80 °C, 100 °C and 120 °C, respectively.

Through modeling and simulation analysis, the results of thermal distribution on the horizontal plane of the computational model were shown in Fig. S3, besides the results of cross-sectional plane shown in Fig. 6d. The plane that the bottom of heating probe located in was chosen as the horizontal plane for analysis. In Fig. S3, the central high temperature zone with uniform color denotes the heating probe. In these three temperature conditions, the cements were continuously heated until their outermost surface reached 60 °C. Apparently, it took different time for the models with three conditions to perform this process.

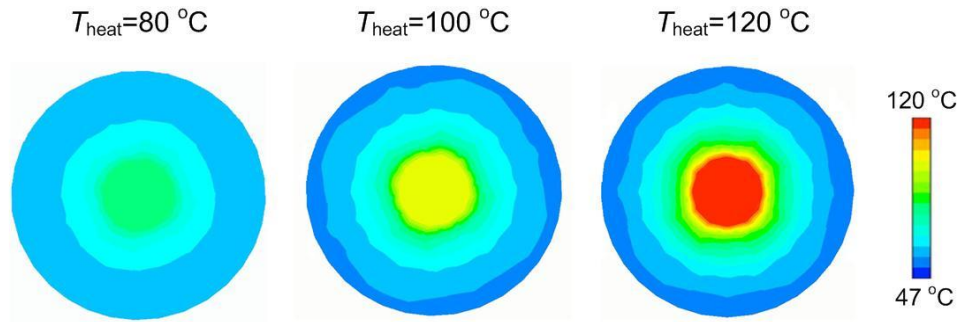

**Figure S3 | The thermal characterization on the horizontal plane of the computational model in the condition of $T_{heat}$= 80 °C, 100 °C, 120 °C.**

**Numerical calculation method**

The software Gambit 2.3 was employed to generate the tetrahedral mesh on theoretical model. The mesh sizes were selected as 1 mm and 3 mm in the domains of heating source and alloy cement, respectively. Accordingly, there were 1821 nodes and 8682 elements generated on two domains. The meshed model was then introduced into the software of Fluent 6.3 for further analysis of the temperature field. The whole simulation was completed by Fluent 6.3 on the Dell PE 2950 workstation with two quad-core CPUs (Intel Xeon x5356 @3.00 Hz) and 8 GB memories.

**Improvement of syringe**

Due to the low melting point of Bi/In/Sn/Zn alloy, the melting alloy will be rapidly solidified without heating at room temperature. To operate the liquid alloy cement with syringe more conveniently, the syringe is improved by introducing an external heater (see Fig. S4). The temperature of the heater is kept above or close to the melting point of alloy cement. We consider that such simple combination will maintain the alloy material always as liquid in the syringe, and it provides higher flexibilities in operation. In addition, other heating methods or devices will be investigated in the near future.

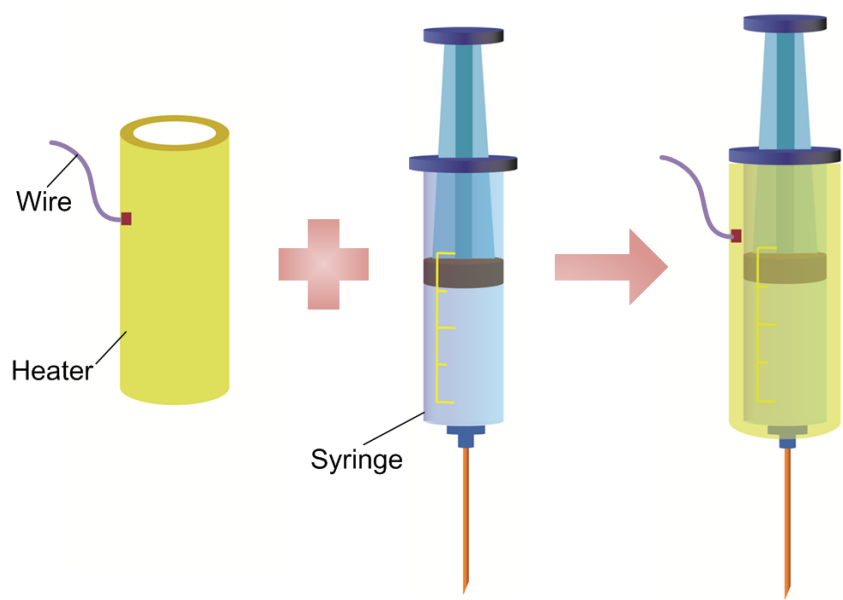

**Figure S4 | The schematic of the heating syringe.** The heating syringe is composed of a heater and syringe. The wire is utilized to connect the electric power providing energy for the heater.

**Acknowledgements:**

We would like to thank L. Wang and H. Ge for the helpful suggestions on materials and Dr. Z. He for the support in model calculation. They are all from Technical Institute of Physics and Chemistry, Chinese Academy of Sciences. We would also like to acknowledge K. Pan (Radiological Department, Hospital of Tsinghua University), H. Du (Tsinghua University), L. Xie (Tsinghua University) and X. Li (Tsinghua University) for the help in operating the instruments.

**Author contributions:**

J. L. conceived, designed the work and wrote part of the manuscript. L. Y. performed all the experiments and wrote the manuscript. C. J. contributed to the theoretical heat transfer calculation and experiment on radiation imaging. All authors discussed the results.